\documentstyle[preprint,aps,12pt]{revtex}

\tightenlines 

\preprint{UT-KOMABA/2000-2, \ 
hep-th/0001057}

\draft

\title{
 $D0-D4$ system and $QCD_{3+1}$
}
\author{ Kenji Suzuki }
\address{
\it Institute of Physics, University of Tokyo, Komaba, \\
\it Tokyo 153, Japan \\
\it ksuzuki@hep1.c.u-tokyo.ac.jp
}

\begin{document}

\maketitle

\begin{abstract}
We consider a $(3+1)$-dimensional QCD model
using a dual supergravity description
with a non-extremal $D0$-$D4$ brane background.
We calculate the spectrum of glueball masses 
and Wilson loops in the background.
The mass spectrum is shown to coincide with one
in non-extremal $D4$-brane systems,
and an area low of spatial Wilson loops is established.
We show that there is a region that Kaluza-Klein modes 
of the Euclidean time direction are decoupled
without decoupling glueball masses.
\end{abstract}


\section{Introduction}

Recently years, Maldacena's conjecture~\cite{mal,IMSY}, 
which is a duality between supersymmetric gauge theory
in the large $N$ strong-coupling limit 
and string theory on 
$AdS$ backgrounds, has been discussed.
For instance, 
the relations between
correlation functions in gauge theory
and effective actions in string theory
on $AdS$ backgrounds
are established~\cite{GKP,wit}.
According to a method discussed in~\cite{mal2}, 
Wilson loops for the large $N$ gauge theories
for the strong coupling limit can be 
obtained by calculating the interaction energy
between  the massive quark and the anti-quark
separated by a distance $L$
in terms of string theory on $AdS$ backgrounds.
The gauge theories are in 
a deconfinement phase.

Witten~\cite{wit2} extended the duality 
to the non-supersymmetric theories at zero temperature, 
which are obtained by the compactification of the Euclidean time direction 
with anti-periodic boundary conditions to fermions.
The compactification radius $R$ is given by $ T=1/2\pi R $,
where $T$ is the Hawking temperature 
in the supergravity description.
The order of the fermion masses is $T$, 
and then supersymmetries are broken. 
%
In non-extremal $D$-brane backgrounds with these boundary 
conditions,
the spectra of glueball masses have been calculated 
by solving the wave equations 
of the dilatons in supergravity description~\cite{RY,BISY,BISY2,GO}.
We observe an area law for spatial Wilson loops, 
which indicates that the large $N$ gauge theories 
at zero temperature
are in a confining phase as expected.
The mass spectra are 
agreement with the lattice 
calculations~\cite{COOT,ORT,mina}.
However, there are some problems that 
Kaluza-Klein states cannot be decoupled 
without decoupling glueball masses in these models.
In order to avoid the difficulty,
the QCD models 
using rotating $D$-brane backgrounds
have been discussed~\cite{rus,CORT,KLT,RS,CRST}.
In these backgrounds, 
KK modes in the Euclidean time direction 
are decoupled without decoupling glueball masses 
in the limit that the angular momenta are infinite.
There are still some unwanted KK states that do not decouple.

In this paper we study a $(3+1)$-dimensional QCD model 
using the supergravity description 
of non-extremal $D0$-$D4$ branes,
in order to obtain the QCD model without KK-modes.
The composition of intersecting $D$-branes 
in supergravity is discussed 
in~\cite{PT,Tsey,GKT,CT,RT,CY,BREBS,BREBS2}.
We calculate the Wilson loop according to the method~\cite{mal2}, 
and the glueball mass spectrum obtained by
solving the wave equation of the dilaton 
in the background.
In the case that 
the dilaton $\phi$ in the limit $r \to \infty$ is finite,
we can not obtain the QCD model 
without KK modes.
However, 
we show that in the case that
the dilaton in the limit $r \to \infty$ vanishes,
there is a region 
where KK modes of the Euclidean time direction are decoupled
without decoupling glueball masses.
Furthermore,
the spatial Wilson loop exhibits a confining
area law behavior, and the glueball mass spectrum is 
coincident with one in the non-extremal 
$D4$-brane background.

The organization of this paper is as follows.
In section 2, we calculate glueball masses
using the supergravity approach with
non-extremal $D0$-$D4$ brane background.
In section 3, we consider the spatial Wilson loops
and compare glueball masses and Kaluza-Klein masses.
In section 4, we consider in the case of
the background that in the limit $r \to \infty$
the dilaton vanishes. 

\section{glueball masses}

We consider glueball masses
using the supergraivity description, 
which are obtained by solving the wave equation of the dilaton 
in the supergravity background.
We treat the background 
corresponding to 
the non-extremal $D0$-$D4$ branes given by
\begin{eqnarray}
ds^2=f(r)^{-1/2}g(r)^{-1/2} \bigg[ -h(r) dt^2
 +g(r)\{dx_1^2+dx_2^2+dx_3^2+dx_4^2\} \nonumber \\
 +f(r)g(r)\{h(r)^{-1}dr^2+r^2d\Omega^2\}\bigg],
\end{eqnarray}
where
\begin{eqnarray}
 f(r)=1+\frac{g\alpha'Q_1}{r^3}, 
 \quad g(r)=1+\frac{g\alpha'Q_2}{vr^3}, 
 \quad h(r)=1-\frac{r_0^3}{r^3},
\end{eqnarray}
with a dilaton background 
\begin{eqnarray}
 e^{-2\phi(r)} = f(r)^{1/2}g(r)^{-3/2}
  =  \sqrt{\frac{v^3r^6(r^3+g\alpha'Q_1)}{(vr^3+g\alpha'Q_2)^3}},
\end{eqnarray}
which is finite in the limit $r \to \infty$.
We consider the metric with the Euclidean time coordinate
and $x_1 \to i x_1$,
and we take the limit 
\begin{eqnarray}
 U = \frac{r}{\alpha'} = fixed , \quad \alpha' \to 0.
\end{eqnarray}
The wave equation of the dilaton is 
\begin{eqnarray}
 \partial_\mu e^{-2\phi}
\sqrt{g}g^{\mu\nu} \partial_\nu\phi=0.
\end{eqnarray}
Assuming that $\phi=e^{ikx_1}\rho(U)$, the wave equation reduces to
\begin{eqnarray}
&& \partial_U(U^3-U_T^3)U \partial_U \rho + M^2 f(U)U^4 \rho
=0 ,
\end{eqnarray}
where $M^2 = -k^2$. 
We denote that the equation is independent of the function
$g(U)$,
and the wave equation in this background 
is coincident with one in the non-extremal $D4$-brane background.
In the limit $ \alpha \to 0 $, the equation is  
\begin{eqnarray}
&& \partial_U(U^3-U_T^3)U \partial_U \rho + M^2 gQ_1 U \rho 
= 0,
\end{eqnarray}
with $U_T=r_0/\alpha'$.
Defining a new variable $x=U^2$ and rescaling, this equation
reduces further to
\begin{eqnarray}
 \partial_x(x^{2+1/2}-x)\partial_x\rho + \sigma \rho = 0,
\end{eqnarray}
with $\sigma = M^2 gQ_1/4U_T$.
Using the WKB approximation~\cite{mina}, the mass spectrum is
\begin{eqnarray}
  M^2 = 16\pi^3 
   \frac{\Gamma(\frac{2}{3})}{\Gamma(\frac{1}{6})}
   (\frac{3}{8\pi})^2 \frac{4U_T}{gQ_1} m(m+2) + O(m^0) .
\quad (m=1,2,3,\cdots)
\end{eqnarray}
If we take into account the dilaton fluctuations~\cite{GKT2}, 
there are some corrections due to the $D0$-branes.

\section{Wilson loop}

We consider the Wilson loop for the full $D0$-$D4$ brane background
discussed in the previous section,
in the region of the small but nonzero $\alpha'$, 
because we need the small curvature 
in the regions of the large but finite $Q_1,Q_2$.
According to the method~\cite{mal2},
the expectation value of the Wilson loop is 
\begin{eqnarray}
 <W(C)> \sim e^{-TE(L)} \sim e^{-S}, 
\end{eqnarray}
where $S$ is the Nambu-Goto action
of a fundamental string.
$C$ denotes a closed loop in the $x_1-x_2$ directions.
$L$ is the distance between the quark and the antiquark.

We consider the spatial Wilson loop in the 
$x_1-x_2$ directions with Euclidean time coordinate
and $x_1 \to -ix_1$.
We take the Euclidean time coordinate
as the space-like circle with the radius $R = 1/2\pi T$,
where $T$ is the Hawking temperature in the
supergravity description.
The fermions obey antiperiodic boundary conditions. 
In the low energy, 
we can obtain the zero temperature $(3+1)$-dimensional 
QCD model.
The world-sheet action in the $x_1-x_2$ directions is 
\begin{eqnarray}
 S_{NG} &=& 
  \frac{1}{2\pi \alpha'} \int dx_1dx_2
  \sqrt{det G_{MN}\partial_\alpha X^M \partial_\beta X^N}
 \nonumber \\
  &=&
  \frac{Y}{2\pi \alpha'} \int dy 
  \sqrt{g(U)h(U)^{-1}(\frac{dU}{dy})^2+g(U)f(U)^{-1}}
 \nonumber \\
  &=& \frac{Y}{2\pi \alpha'} \int dy
  \sqrt{\frac{\alpha'^2U^3v+gQ_2}{(U^3-U_T^3)v}
 (\frac{dU}{dy})^2
   +\frac{\alpha'^2U^3v+gQ_2}
 {(\alpha'^2U^3+gQ_1)v}},
\end{eqnarray}
with $U_T =  r_0/\alpha'$, and  $y\equiv x_2$.
$Y$ is a period in the $x_1$-direction.
$G_{MN}$ is the string metric of the non-extremal 
$D0$-$D4$ brane background discussed in the previous section.
%
The distance between the quark and the antiquark is  
\begin{eqnarray}
 L = 2 \int dy 
 &=& 
 2 \sqrt{g(U_0)/f(U_0)} \int_{U_0}^\infty dU
   \sqrt{\frac{g(U)f(U)}{g(U)h(U)(g(U)/f(U)-g(U_0)/f(U_0))}}
\nonumber \\
 &=& 
 2 \frac{1}{\alpha'\sqrt{gQ_1v-gQ_2}}
   \int_{U_0}^\infty dU
   \sqrt{\frac
{(\alpha'^2U^3+g Q_1)^2(\alpha'^2 U_0^3 v + gQ_2)}
{(U^3-U_T^3)(U^3-U_0^3)}}
\nonumber \\
 &=& 
 2 \frac{1}{\alpha'\sqrt{gQ_1v-gQ_2}U_0^4}
   \int_{1}^\infty dx
   \sqrt{\frac
{(\alpha'^2x^3U_0^3+g Q_1)^2(\alpha'^2 U_0^3 v + gQ_2)}
{(x^3-\lambda^3)(x^3-1)}} \\
&\to& 2 \frac{g Q_1\sqrt{U_0^3 v + gq_2}}{\sqrt{gQ_1v-gQ_2}U_0^4}
   \int_{1}^\infty dx
   \frac{1}
{\sqrt{(x^3-\lambda^3)(x^3-1)}} , \quad (\alpha' \to 0)
\end{eqnarray}
where $x=U/U_0$ and $\lambda=U_T^3/U_0^3$.
$U_0$ denotes the lowest value of $U$,
and $q_2$ is defined by $Q_2=\alpha'^2q_2$. 
The energy is 
\begin{eqnarray}
 E_{q\bar{q}}
 &=& \frac{1}{\alpha'\pi}\int_{U_0}^\infty dU
   \bigg[
   \sqrt{\frac{g(U)^2/f(U)}{h(U)(g(U)/f(U)-g(U_0)/f(U_0))}} 
 - 1
   \bigg]
 -\frac{1}{\pi}\int_{U_T}^{U_0}dU \sqrt{g(U)/h(U)}
 \nonumber \\
 &=& \frac{1}{\pi\alpha'^2\sqrt{v(gQ_1v-gQ_2)}}
 \int_{U_0}^\infty dU
  \bigg[
 \sqrt{\frac{(\alpha'^2U_0^3+gQ_1)(\alpha'^2U^3v+gQ_2)^2}
 {(U^3-U_T^3)(U^3-U_0^3)}}
 -1 
 \bigg]
 + \frac{U_T-U_0}{\pi}
\nonumber \\
 &=& \frac{1}{\pi\alpha'^2\sqrt{v(gQ_1v-gQ_2)}U_0^4}
 \int_{1}^\infty dx
  \bigg[
 \sqrt{\frac{(\alpha'^2x^3U_0^3+g Q_1)(\alpha'x^3U_0^3v+gQ_2)}
 {(x^3-\lambda^3)(x^3-1)}}
 -1 
 \bigg]
 + \frac{U_T-U_0}{\pi}
\nonumber \\
 &\to& \frac{\sqrt{gQ_1}}{\pi\sqrt{v(gQ_1v-gQ_2)}U_0^4}
 \int_{1}^\infty dx
  \bigg[
 \sqrt{\frac{(x^3U_0^3v+gq_2)^2}
 {(x^3-\lambda^3)(x^3-1)}}
 -1 
 \bigg]
 + \frac{U_T-U_0}{\pi} . \quad (\alpha' \to 0)
\end{eqnarray}
We consider the large $L$ behavior, which is obtained
in the limit $\lambda \to 1$. 
In this limit, the main contribution to the integrals
of $L$ and $E$ comes from the region near $x = 1$.
Therefore we obtain the string tension as
\begin{eqnarray}
 T_{TM} = E_{q\bar{q}}/L 
   &=& \frac{\sqrt{\alpha'^2U_0^3v+gQ_2}}
         {2\alpha'\pi \sqrt{\alpha'^2U_0^3v+gQ_1v}}
  \\
     &\to& \sqrt{\frac{U_T^3}{4\pi^2 gQ_1}} 
  \quad  (gq_2<< U_0^3v)
  \\
     &\to& \sqrt{\frac{q_2}{4\pi^2Q_1v}}. \quad (gq_2 >>U_0^3v)
\end{eqnarray}

The glueball mass is $M_{GB} \sim \sqrt{U_T/gQ_1}$,
as discussed in the previous section.
Kaluza-Klein masses are proportional to $1/R$,
where $R=1/T$ is the compactification
radius of the time coordinate.
The Hawking temperature $T$ in the supergravity description
is given by
\begin{eqnarray}
T &=& \frac{3\alpha'U_T^2\sqrt{v}}
{4\pi\sqrt{(\alpha'^2U_T^3+gQ_1)(\alpha'^2U_T^3v+gQ_2)}}
  \\
     &\to& \frac{3\sqrt{U_T}}{4\pi\sqrt{gQ_1}} 
  \quad  (gq_2<< U_0^3v)
  \\
     &\to& \frac{3U_T^2\sqrt{v}}{4\pi g\sqrt{Q_1q_2}} .
  \quad (gq_2 >>U_0^3v) 
\end{eqnarray}
In the limit $\alpha' \to 0$,
  KK masses are 
$M_{KK} \sim U_T^2\sqrt{v}/\sqrt{gQ_1(U_0^3v+gq_2)}$.
Then the ratio of the masses is
\begin{eqnarray}
 M_{KK}/M_{GB} &=& \sqrt{U_T^3v/(U_T^3v+gq_2)} 
     \label{ratio}
 \\
 &\to& 1 \quad \quad (gq_2 << U_T^3v)  \\
 &\to& \sqrt{U_T^3/gq_2}<< 1. \quad (gq_2 >> U_T^3v)
\end{eqnarray}
Therefore we can not obtain the region that 
KK masses in the Euclidean time direction
are decoulped without decoupling 
glueball masses.
We consider the way to resolve this problem in the next section.

\section{background with zero dilaton at boundary}

We consider the boundary of 
the non-extremal $D0$-$D4$ brane background
discussed in previous section.
The background is the Minkowski space at $r \to \infty$, 
and there is no boundary.
In the supergravity - SYM correspondence~\cite{IMSY}, 
it is needed that the background has a boundary at $r \to \infty$.
In addition, we need a QCD model that KK masses are decoupled without 
decoupling glueball masses. 
In order to resolve their problems, 
we replace the harmonic function with
\begin{eqnarray}
 g(r)=1+\frac{g\alpha'Q_2}{vr^3} 
\to g'(r) = \frac{g\alpha'Q_2}{vr^3},
\end{eqnarray}
which corresponds to 
\begin{eqnarray}
 U_T^3v+gq_2 \to gq_2 ,
\end{eqnarray}
in the equation 
of the ratio between glueball masses and KK masses (\ref{ratio}).
Then we can take the region $M_{KK}/M_{GB} >> 1$.
We note that this procedure is not to take the near horizon limit, 
but to replace the harmonic function
in order to obtain the zero dilaton at the boundary.
This means that the effects of the dilaton for Kaluza-Klein
masses in the Euclidean time direction are suppressed
at the boundary.
Then we consider the metric given by
\begin{eqnarray}
ds^2=f(r)^{-1/2}g'(r)^{-1/2}\bigg[-h(r) dt^2
  +g'(r)\{dx_1^2+dx_2^2+dx_3^2+dx_4^2\} \nonumber \\
  +f(r)g'(r)\{h(r)^{-1}dr^2+r^2d\Omega^2\}\bigg],
\end{eqnarray}
where
\begin{eqnarray}
 f(r)=1+\frac{g\alpha'Q_1}{r^3}, 
 \quad g'(r)=\frac{g\alpha'Q_2}{vr^3}, 
 \quad h(r)=1-\frac{r_0^3}{r^3},
\end{eqnarray}
and the dilaton is 
\begin{eqnarray}
 e^{2\phi(r)} 
  =  \sqrt{\frac{(g\alpha'Q_2)^3}{v^3r^6(r^3+g\alpha'Q_1)}},
\end{eqnarray}
which vanishes in the limit $r \to \infty$.
The temperature is
\begin{eqnarray}
 T \sim U_T^2/\sqrt{g^2Q_1q_2}. \quad (\alpha'^2U_0^3 << gQ_1)
\end{eqnarray}
A distance $L$ between the quark and anti-quark is
\begin{eqnarray}
 L &=& 
 2 \frac{1}{\alpha'\sqrt{gQ_1v-gQ_2}U_0^4}
 \int_{1}^\infty dx
 \sqrt{\frac
{(\alpha'^2x^3U_0^3+g Q_1)^2 gQ_2}
{(x^3-\lambda^3)(x^3-1)}}
\nonumber \\
 &\to& 
 2 \frac{gQ_1\sqrt{gq_2}}{\sqrt{gQ_1v-gQ_2}U_0^4}
 \int_{1}^\infty dx
 \frac{1}{\sqrt{(x^3-\lambda^3)(x^3-1)}}, \quad (\alpha' \to 0)
\end{eqnarray}
where $x=U/U_0$ and $\lambda=U_T^3/U_0^3$.
The energy is 
\begin{eqnarray}
 E_{q\bar{q}}
 &=& \frac{1}{\pi\alpha'^2\sqrt{v(gQ_1v-gQ_2)}U_0^4}
 \int_{1}^\infty dx
 \sqrt{\frac{(\alpha'^2U_0^3+gQ_1)(gQ_2)^2}
 {(x^3-\lambda^3)(x^3-1)}}
 + \frac{U_T-U_0}{\pi}
\nonumber \\
 &\to& \frac{gq_2\sqrt{gQ_1}}{\pi \sqrt{v(gQ_1v-gQ_2)}U_0^4}
 \int_{1}^\infty dx
 \frac{1}{\sqrt{(x^3-\lambda^3)(x^3-1)}}
 + \frac{U_T-U_0}{\pi}, \quad (\alpha' \to 0)
\end{eqnarray}
where $U_0$ is the lowest value of $U$.
We denote that 
the mass term of the W-boson  
is removed 
in the background. 
We take the limit $\lambda = U_T^3/U_0^3 \to 1$,
the string tension is
\begin{eqnarray}
 T_{YM} &=& E_{q \bar{q}}/L
 \sim \sqrt{\frac{q_2}{4\pi^2 Q_1}}.
\end{eqnarray}
The string tension is proportional 
to the squared mass of glueballs, namely
\begin{eqnarray}
 T_{YM}=g_{eff}M_{GB}^2,
\end{eqnarray} 
where we define $g_{eff} = \sqrt{g^2Q_1q_2/4\pi^2U_T^2}$ as 
the effective coupling of the theory 
following~\cite{IMSY,BISY2,rus}.
Using this ralation, glueball masses and KK masses
are rewritten by
\begin{eqnarray}
M_{GB}= \sqrt{\frac{U_T}{gQ_1}},
\quad
T \sim \frac{U_T^2}{\sqrt{g^2Q_1q_2}}
 \sim \frac{U_T}{g_{eff}}, 
\end{eqnarray}
and the string tension is rewritten by
\begin{eqnarray}
T_{YM} = \sqrt{\frac{q_2}{4\pi^2Q_1}}
 = g_{eff}\frac{U_T}{gQ_1} \equiv c \ :fixed.
\end{eqnarray}
The curvature is
\begin{eqnarray}
\alpha'R \sim
 \frac{r_0}{\sqrt{g^2Q_1Q_2}}
=\frac{U_T}{\sqrt{g^2Q_1q_2}} \sim \frac{1}{g_{eff}},
\end{eqnarray}
and the dilaton is rewritten by
\begin{eqnarray} 
 e^{\phi}|_{U=U_T} = g_{eff}^{3/2}\frac{\sqrt{\alpha'}}{Q_1}.
\end{eqnarray}
The supergravity solutions describing $D$-branes
can be trusted if the curvature
in string units and the effective string coupling constant are
small. 
Then we need to take the region 
\begin{eqnarray}
1 << g_{eff} << (Q_1^2/\alpha')^{1/3}.
\end{eqnarray}
In addition, in order to decouple Kaluza-Klein masses
in the Euclidean time direction,
we need to take the region 
\begin{eqnarray}
M_{GB} << M_{KK} \sim T.
\end{eqnarray}
Therefore 
we can obtain the QCD model
that KK masses are decoupled without decoupling 
glueball masses
in the region
\begin{eqnarray}
 g(Q_1\alpha')^{1/3}c
 << (gQ_1c^2)^{1/3} 
 <<  U_T << gQ_1 c .
\end{eqnarray}



\section{conclusion}

We have studied the $(3+1)$-dimensional QCD model using the 
supergravity description of the non-extremal $D0$-$D4$ branes 
without KK-modes in the Euclidean time direction.
In the case 
of the supergravity background whose 
dilaton in the limit $r \to \infty$ is finite,
we cound not obtain the QCD model without decoupling KK-modes.
However, in the case of the background whose dilaton 
in the limit $r \to \infty$ vanishes, we have found the region where
KK modes in the Euclidean time direction 
are decoupled
without decoupling glueball masses. 
The background has the boundary at $r \to 0$, 
which is needed in the supergravity - SYM correspondence.
There are still some unwanted KK states
that do not decouple.
We have calculated the Wilson loops 
which exhibit the confining
area law behavior, and the glueball mass spectrum
is coincident with one 
in the non-extremal $D4$-brane background.
If we take into account dilaton fluctuations~\cite{GKT2}, 
there are some corrections due to the $D0$-branes.
We need to study the physical reasoning 
of the replacement of the harmonic function. 

Comparing our results with that using rotating
$D$-branes, discussed in~\cite{rus,CORT,KLT,RS,CRST},
we share the results that the Yang-Mills tension is finite,
and glueball masses are finite and small
in the region that the curvature is small. 
The calculations of glueball masses and 
the string tension using the Wilson loop are
more tractable in the our metric than that 
in the rotating $D$-branes.
This is expected to be useful for further applications
of our approach.
For instance,
Wilson loops of the baryon may be explicitly computed using 
our metric.

\acknowledgements 
I am grateful to Professor  
T. Yoneya
for discussions and 
for giving me valuable advice.



\end{document}